# Complete suppression of flux instabilities in ramped superconducting magnets with synchronous temperature-modulated $J_c$

Cun Xue[1], Han-Xi Ren[2], Kai-Wei Cao[3], Wei Liu[4,5], Wen-Tao Zhang[4,5], Fang Yang[1],

Guo Yan[4], You-He Zhou[6], Pingxiang Zhang[1,7]

*1 Institute of Superconducting Materials and Applied Technology, Northwestern Polytechnical University, Xi'an 710072, China*

*2 School of Aeronautics, Northwestern Polytechnical University, 710072 Xi'an, China*

*3 School of Mechanics, Civil Engineering and Architecture, Northwestern Polytechnical University, 710072 Xi'an, China*

*4 Western Superconducting Technologies Co. Ltd, Xi'an 710018, People's Republic of China*

*5 Xi'an Superconducting Magnet Technology Co. Ltd, 710014 Xi'an, China*

*6 Department of Mechanics and Engineering Sciences, Lanzhou University, Lanzhou 730000, China*

*7 Northwest Institute for Nonferrous Metal Research, Xi'an 710016, People's Republic of China*

**Abstract**

Nonlinear multi-field coupling as an intrinsic property of complex physical systems often leads to abrupt and undesired instabilities. For current-ramped high-field $Nb_3Sn$ magnets, frequent flux jumps are observed, which easily causes premature quenches and requires prolonged and resource-intensive magnet training process. In this study, we propose a paradigm-shifting methodology framework that achieves complete suppression of thermomagnetic instabilities through synchronized temperature-modulated critical current density ($J_c$). Through numerical simulations of flux jumps in multifilamentary $Nb_3Sn$ wires at various temperatures, we construct thermomagnetic stability diagram in the $H_a$–$T$ plane. The simulated results are in good agreement with experiments, confirming that the synchronized temperature ramp-down can fully eliminate flux jumps. We reveal the underlying mechanism of enhancing the thermomagnetic stability arises from that synchronized temperature ramp-down can continuously tune both $J_c$ and its slope. Furthermore, we explore the thermomagnetic instabilities of current-ramped superconducting magnets through large-scale GPU-optimized algorithm. The flux jump and quench diagram in the $I_a$–$T$ plane are obtained. It indicates that the temperature ramp-down can completely suppress flux jumps without compromising $J_c$ at high magnetic fields. Importantly, this method does not require modifications to the superconducting microstructures or fabrication process, offering a practical and broadly applicable solution. The findings not only provide a robust method for stabilizing various superconducting magnet systems, including high-temperature superconducting magnets wound with second-generated (2G) coated tapes, but also suggest a generalizable strategy for controlling instability in other nonlinear non-equilibrium physical systems.

The natural world is rife with complex physical systems where nonlinear couplings between multiple fields give rise to sudden, often catastrophic transitions under external perturbations. From the fragmentation of brittle solids and the initiation of avalanches to electrical breakdown in storms, these emergent instabilities reflect universal challenges in controlling coupled energy landscapes across vastly different length and time scales[1-5]. For type-II superconductors, such as $Nb_3Sn$ exposed to a time-varying magnetic field, magnetic vortices driven by Lorentz forces dissipate energy through local Joule heating, triggering rapid flux redistribution and thermal runaway—a phenomenon known as thermomagnetic instability of flux jump [6-9]. These abrupt flux jumps in $Nb_3Sn$ superconductors pose a critical operational challenge, inducing premature quenches at fields and currents far below engineering critical thresholds [10-12]. This instability requires prolonged and resource-intensive magnet training protocols to achieve target operational parameters, significantly escalating cryogenic costs and system downtime [13, 14]. Moreover, the stochastic nature of flux jumps introduces spatial and temporal field inhomogeneities within the magnet bore, thereby complicating precision field-correction algorithms that are essential for applications demanding sub-ppm stability, such as particle beam steering [15-17]. Attempts to mitigate such events using quench detection systems based on voltage transients have also proven unreliable, as flux jumps may trigger spurious signals indistinguishable from true quench precursors [15, 18]. Consequently, the flux jumping remains a critical bottleneck for applications in next-generation accelerators, fusion reactors, and high-field medical imaging.

Over the past decades, various approaches have been explored to suppress flux jumps in $Nb_3Sn$ superconductors, with an emphasis on optimizing wire structure and metallographic structure of the material. Studies from Fermilab [19] and Brookhaven National Laboratory [20, 21] have demonstrated that reducing filament size and increasing the residual resistivity ratio (RRR) of the copper matrix can improve low-field stability, although these strategies often introduce trade-offs in manufacturability and critical current density ($J_c$) [22]. Grain refinement and the introduction of artificial pinning centers (APCs) via internal oxidation have shifted the pinning force peaks to higher fields, substantially enhancing high-field $J_c$ [23-28]. Nevertheless, these methods fall short of completely eliminating low-field flux jumping. Other strategies, such as high-temperature heat treatment and Ti doping [29], have improved the upper critical field ($B_{c2}$) and specific heat, yet at the cost of degraded high-field performance. For bulk $Nb_3Sn$, modest Sn enrichment [30], grinding, and sintering have yielded notable gains in low-field stability, although often accompanied by undesired reductions in $J_c$ across the field range.

Enhancing the specific heat of superconducting conductors has emerged as a promising alternative to suppress flux jumps. Various approaches have been investigated, including the embedding of low-temperature high-specific-heat (LHCS) materials into epoxy matrices [31-42], hybrid cabling techniques, and direct LHCS doping into superconducting strands. While embedding techniques yielded limited MQE improvements due to poor thermal conductivity, direct doping offered dramatic stabilization at the expense of mechanical integrity. Recent advances by Fermilab achieved twofold MQE enhancement in 50-meter $Nb_3Sn$ conductors [43, 44], although narrow processing margins and mechanical challenges persist. Attempts to further improve filament thermal conductivity and mechanical robustness [44-46] have yielded promising short-sample results, but scalable industrial fabrication remains elusive.

In this work, we present a breakthrough of new approach to suppress flux jump in $Nb_3Sn$ high-field magnets by synchronously controlling the dynamic temperature during the current ramp process. Through a precise real-time temperature modulation strategy, we achieve full suppression of flux jumps without sacrificing $J_c$ or other properties. This dynamic temperature-modulated methodology paves a new way to enhance the flux instability of the next generation of high-field superconducting systems.

**Fully Suppression of Flux Jumps in Superconducting Wires**

In a first step, we will unveil the thermomagnetic instabilities of multifilamentary high-$J_c$ $Nb_3Sn$ wire at different working temperatures. In this case, we carry out magnetization measurements for a short sample from Oxford Superconducting Technology (OST). As shown in Fig. 1(a), the sample with length of 4.9 mm is prepared with polishing two ends before experimental measurements (see details of the sample in Ref. [47]). Fig. 1(b) represents the experimental magnetization loops of the sample with field-ramping rate of 10 mT/s at different working temperatures. Note that we only show the branches with positive magnetic field in order to see the details clearly. It is apparent that the number of the spikes in the magnetization curve induced by flux jumps decreases with increasing working temperature. It indicates that higher working temperature can significantly suppress the flux jumps of superconductors. Particularly, the flux jumps of signal bare superconducting wire completely vanish when the working temperature is increased up to 7 K.

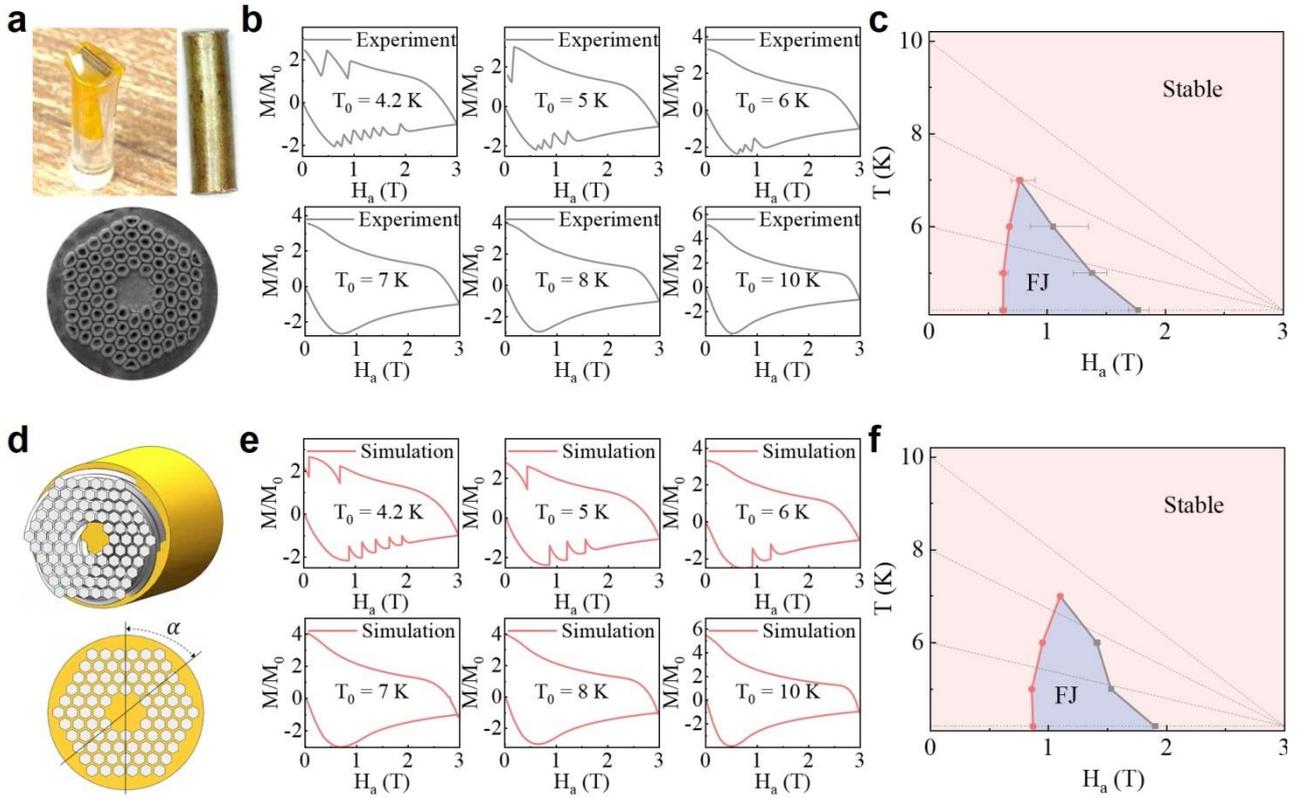

Fig. 1. (a) The short sample and SEM image of its cross-section used in magnetization experiments. (b) The magnetization loop of the short sample with field-ramping rate of 10 mT/s at different working temperatures. (c) The thermomagnetic stability diagram in the $H_a$-$T$ plane obtained through the statistics of experiments. (d) The schematic of the numerical model for the multifilamentary wire. (e) The simulated magnetizations of the multifilamentary wire at the same parameters with experiments. (f) The simulated thermomagnetic stability diagram in the $H_a$-$T$ plane.

The cross-section of the multifilamentary wire does not exhibit continuous rotational symmetry. In this case, the magnetic field threshold of the flux jumps on the superconducting wires should exhibit angular dependence relative to the magnetic field orientation. To obtain statistical results, multiple magnetization experiments with mounting and remounting samples were conducted at each working temperature (see all experimental results in Supplementary Material). Fig. 1(c) shows the thermomagnetic stability diagram in the $H_a$-$T$ plane obtained through the statistics of experimental data for field ramping-rate of 10 mT/s. One can see the region of flux jumps are determined by two critical parameters. One is the onset of thermomagnetic instabilities (Red curves), representing the superconducting systems accumulate sufficient electromagnetic energy to trigger the flux jumps. Another one is the critical magnetic field of terminating flux jump events in the superconducting system (Grey curves). With increasing applied magnetic field under a specific temperature, one can observe flux jumps in multifilamentary superconducting wires if the electromagnetic energy is sufficient to trigger the thermomagnetic instabilities prior to reaching the threshold of terminating flux jumps; otherwise, the flux jumps cannot be detected.

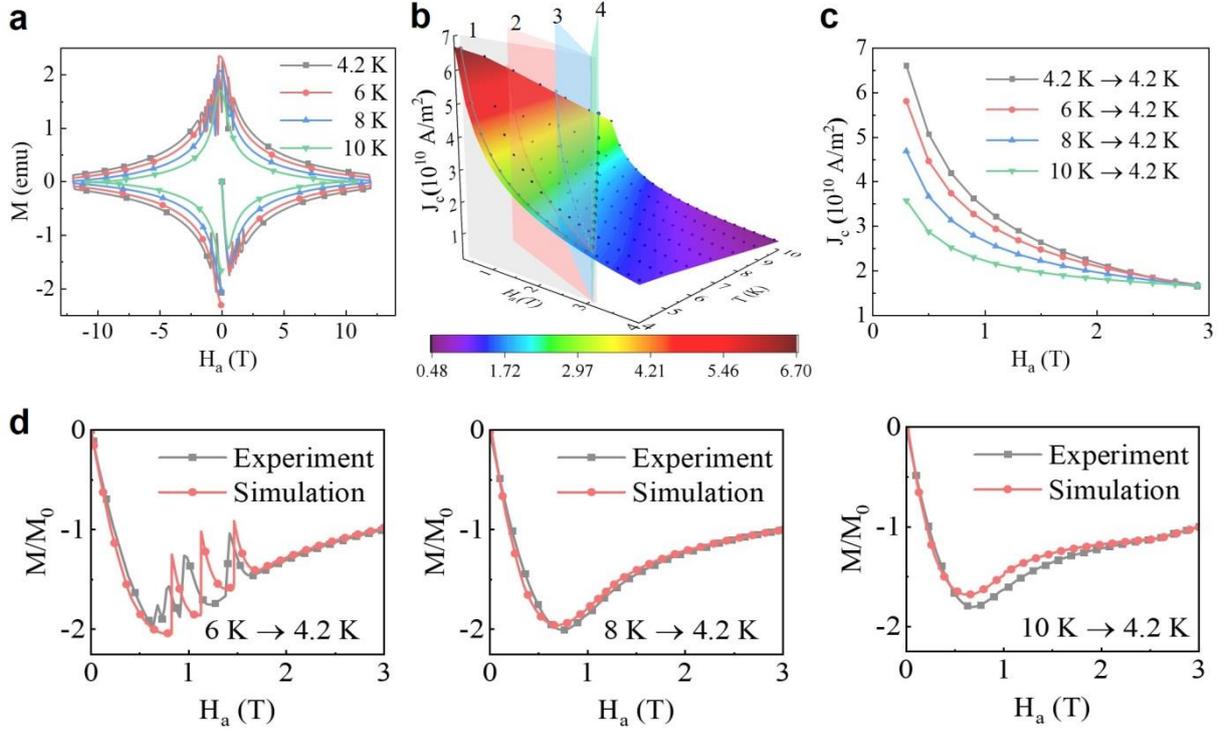

Fig. 2. (a) The experimental magnetization loops of the short multifimentary wire by sweep magnetic field within ±12 T at different working temperatures. (b) The critical current density $J_c$ as a function of magnetic field and temperature. (c) The variations of critical current $J_c$ with magnetic field along four different temperature routes depicted in (b). (d) The experimental and simulated magnetizations of multifilamentary wires along temperature ramp-down routes during the field ramp process.

To elucidate the flux jumps observed in the multifilamentary superconducting wire and the relevant coils discussed later, we developed numerical framework with A-V method by coupling with the heat conduction equation. The computational framework was implemented with home-made codes using the CUDA parallel computing platform and optimized for execution on GPU architectures to leverage massive parallel processing capabilities (see details in Ref. [47]). The computational methodology has been verified through benchmark simulations comparing with a serious of experiments. In this work, utilizing the numerical model shown in Fig. 1(d) along with the accurate $J_c$ as a function of magnetic field and temperature (see detailed discussions in the following text), we obtain the simulated magnetization of the multifilamentary wire at various working temperatures (see Fig. 1(e)). In order to consider the effects of magnetic field orientations, a statistical averaged result of the thermomagnetic stability diagram in the $H_a$-$T$ plane as shown in Fig. 1(f) can be obtained based on the numerical cases with different angles of magnetic field ($\alpha$=0, 10°, 20°, 30°, see details in Supplementary Material). The main characteristics of the simulated magnetization loop and the flux jumps diagram in the $H_a$-$T$ plane can reproduces the experimental results well. The observed minor quantitative discrepancies between experiments and simulations should be attributed to irregular filament geometries and the presence of tin-cores localized within the filamentary structures.

Both the experiments and numerical simulations demonstrate that the critical field of terminating flux jumps progressively diminishes with increasing temperature. It is attributed to the critical current density as a function of temperature and magnetic field, i.e., $J_c(B, T)$. By carrying out more magnetization experiments with sweeping magnetic field within ±12 T at different working temperatures (see Fig. 2(a)), we could obtain an accurate $J_c(B, T)$ as shown in Fig. 2(b) (see details in Supplementary Material and Ref. [47]). With increasing working temperature, we can see that the absolute value of $J_c$ and its slope with magnetic field significantly decreases, which leads to the disappearance of flux jumps under higher working temperature. While elevated operating temperatures can effectively suppress flux jumps, the accompanying rapid degradation of critical current density in superconducting wires at high temperatures inevitably induces premature quenches prior to achieving the designed magnetic field targets. Consequently, this thermal suppression strategy with constant high temperature cannot be viably implemented in practical superconducting magnet.

To address this challenge, we can select other routes of temperature from initial 6 K, 8 K and 10 K to terminal 4.2 K marked with routes 2, 3 and 4 as shown in Fig. 2(b), which representing a simultaneous progressive reduction in working temperature during magnetic field ramp process. Fig. 2(c) demonstrates that the slope of $J_c$ can be significantly decreased with the temperature ramp-down approach. Additionally, the temperature routes 2-4 are also plotted with dashed lines in Fig. 1(c, f). We suppose this temperature ramp-down strategy can simultaneously circumvent flux jump regimes while preserve the critical current density of superconducting wires under high magnetic fields. For the purpose of validation, we conduct additional magnetization experiments and performed numerical simulations on superconducting wires with simultaneously increasing magnetic field and varying temperature. As shown in Fig. 2(d), the temperature ramp-down strategy can indeed suppress the flux jumps during the field ramp process. There are no any spikes observed in the magnetization and the flux jump instability is entirely eliminated with the temperature routes 10→4.2 K and 8→4.2 K (see more detailed experimental and simulated results in Supplementary Information).

**Suppression of Flux Jumps in Superconducting coils**

As shown in Fig. 3(a), high-field superconducting magnets are fabricated through densely wound configurations of high-$J_c$ multifilamentary superconducting wires with structural stabilization achieved via epoxy resin impregnation and curing protocols. As demonstrated in Ref. [47], $Nb_3Sn$ magnet systems exhibit fundamentally distinct flux jump dynamics compared to a single bare $Nb_3Sn$ wire. This complexity arises from two synergistic mechanisms: (i) spatially heterogeneous magnetic field ramping-rates across superconducting wires, (ii) strong thermal coupling between neighboring wires. Crucially, thermal perturbations generated by localized flux jumps propagate through the magnet's thermal network, thereby inducing premature thermomagnetic instabilities in adjacent wires below their intrinsic critical thresholds. Consequently, the global instability of superconducting magnets manifests not as linear superposition of individual wire behaviors, but rather as nonlinear spatiotemporal propagation of coupled thermomagnetic disturbances. In this case, we further explore the temperature-modulated suppression of flux jumps in the superconducting coils.

Our numerical model consists a solenoid coil configuration with $N_x \times N_y$ winding turns, subjected to linearly ramped transport current $I_a(t)$ as illustrated in Fig. 3(a). The field-dependent critical current density ($J_c$) characteristics of the multifilamentary superconducting wire at various operational temperatures are presented in Fig. 3(b). Both 40×40 and 20×20 turn configurations exhibit linear scaling between peak self-field generation and applied current density. The quench current for these coil geometries under different fixed temperatures are determined by critical intersection coordinates between the transport current load line and $J_c(B,T)$ characteristics, visualized as demarcation curves separating stable operation from quench regimes in Fig. 3(c, d).

Utilizing the large-scale GPU-optimized numerical algorithm, we systematically conduct electromagnetic-thermal coupling simulations at different fixed working temperatures in order to map the thermomagnetic instability boundaries in temperature-applied current ($I_a$-$T$) plane for both coil configurations (cases of 40×40 and 20×20 as shown in Fig. 3(c, d)). Remarkably, persistent flux jumps are observed at 4.2 K until transport current exceeds 600 A for the coil with 40×40 turns. This phenomenon arises from the localized field in the inner region of the coil is still in the regime of thermomagnetic instabilities as discussed in Ref. [47].

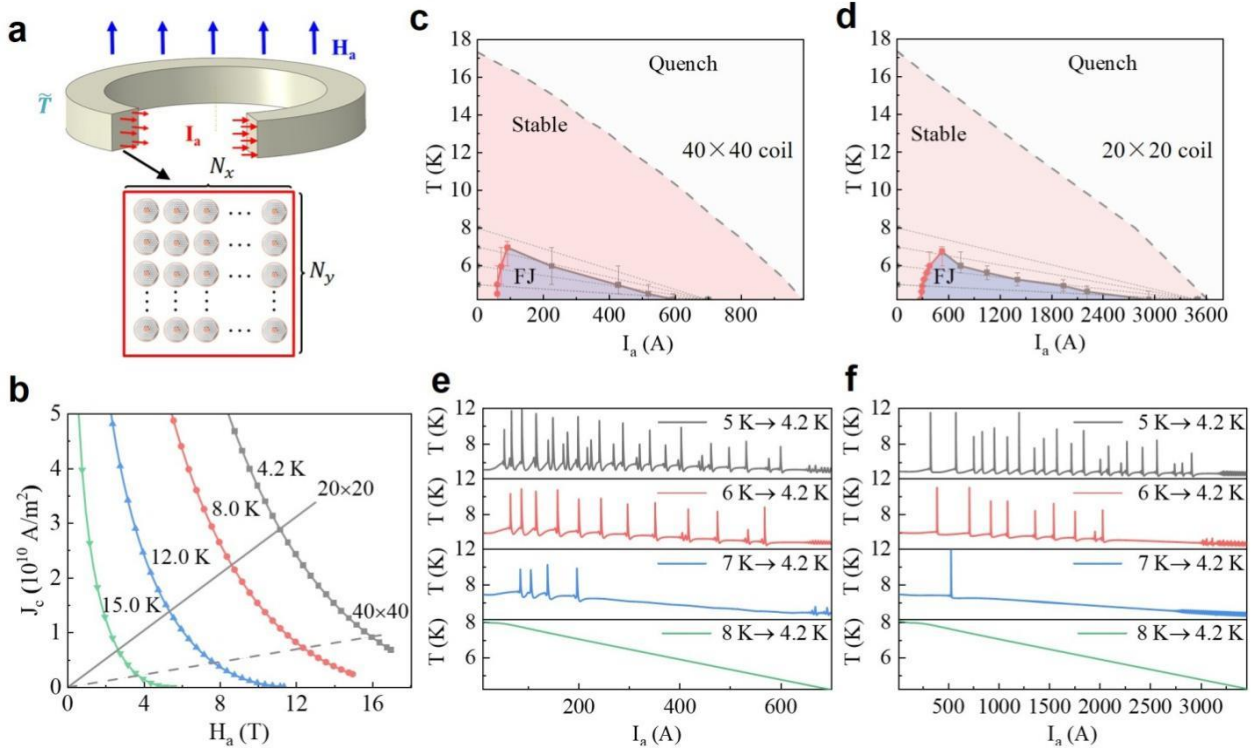

Fig. 3. Schematic of numerical model for a solenoid superconducting coil with $N_x \times N_y$ winding turns. (b) The variations of critical current density with magnetic field at different working temperatures. The ramped transport current for the coils with 40×40 and 20×20 turns. (c, d) Thermomagnetic stability diagram and the quench current trajectories in the temperature-applied current ($I_a$-$T$) plane for the coils with 40×40 and 20×20 turns with current ramping-rate of 1 A/s. (e, f) The time-evolutions of maximum temperature in the coils during current ramp-up with progressive cooling from 8 K, 7 K, 6 K, and 5 K to a terminal state of 4.2 K.

Having established the thermomagnetic stability diagram and the quench current trajectories in the $I_a$-$T$ plane for the studied coil configuration, four distinct thermal control protocols are implemented during current ramp-up process: progressive cooling from initial working temperatures of 8 K, 7 K, 6 K, and 5 K to a terminal state of 4.2 K (depicted as dashed trajectories in Fig. 3(c, d)). Numerical modeling results for the 40×40 turn configuration, presented in Fig. 3(e), reveal fundamentally different time-evolution of maximum temperature in the coil. The 8→4.2 K cooling trajectory demonstrates exceptional thermal stability, exhibiting smooth temperature progression devoid of characteristic flux jump spikes and conclusive evidence of complete flux jump suppression. This suppression mechanism is well consistent with the protocol's strategic avoidance of the identified flux jump regime in the phase diagram (see Fig. 3(c, d)). Furthermore, the flux jump region in the $I_a$-$T$ phase diagram exhibits dependence not only on the number of turns in superconducting coils but also on the current ramping-rate (see Fig. S7 in Supplementary Material). Additionally, it is worth noting that modifications in coil configurations, such as racetrack coils, will correspondingly alter the flux jump regions, which is needed to explore in further work.

To facilitate engineering applications, we introduce the following generalized strategy for circumventing flux jumps in superconducting magnet systems. Considering a magnet designed for operation at temperature $T_0$ (e.g., Nb$_3$Sn magnets typically operating at 4.2 K), the temperature trajectory should be initiated at 10–12 K followed by linear cooling synchronized with the current ramp process. This approach leverages the characteristic convex temperature dependence of the magnet's quench current curve. The terminal condition requires the cooldown path to reach 4.2 K with the operating current maintained at 0.8–0.9 times the quench current, i.e., $0.8I_c(T_0)$ or $0.9I_c(T_0)$. This dual-constrained protocol achieves concurrent mitigation of flux jump phenomena while preserving a sufficient safety margin against quench initiation, as the coordinated temperature-current pathway systematically avoids both instability regimes.

To systematically validate the theoretical framework, it is imperative to implement complementary experimental investigations. We propose two feasible experimental schemes as shown in Fig. S8 in Supplementary Material. (i) Active Thermal Control Scheme: Integration of heating resistors external to the superconducting coil enables precise temperature variations through current-controlled Joule heating. This permits programmable cooling trajectories synchronized with energization processes. (ii) Coaxial Current Injection Scheme: A modified conductor design employing coaxial cable geometry could be implemented, incorporating an insulated thin copper core along the central axis. By applying an independent current through this copper channel – decoupled from the main transport current – localized resistive heating can be achieved while maintaining electrical isolation, thereby enabling in-situ temperature modulation during magnet operation.

In summary, this work presents a synchronous temperature-modulated strategy to comprehensively suppress thermomagnetic instabilities in ramped superconducting magnets. Systematic investigations were conducted through numerical simulations and experimental magnetization measurements on a single multifilamentary wire, establishing a thermomagnetic stability phase diagram in the magnetic field-temperature ($H_a$-$T$) plane. We demonstrate that complete suppression of flux jumps can be achieved through real-time temperature-modulated critical current density. Furthermore, large-scale GPU-accelerated parallel computations were implemented through home-made CUDA-coded programs to simulate solenoid magnet stability, reconstructing the thermomagnetic stability phase diagram in current-temperature ($I_a$-$T$) plane. Numerical results confirm that synchronized cooling during current ramp process not only fully suppresses flux jumps but also maintains operational currents below the quench threshold, thereby achieving dual stability against both flux jumps and quench.

This paradigm-shifting approach addresses the challenges of intrinsic flux jump suppression in high-field $Nb_3Sn$ magnets. The proposed methodology offers significant advantages by avoiding the necessity for altering the superconducting microstructure or modifying established fabrication processes of superconducting wires, while simultaneously maintaining the quench critical current of the magnet system. The findings presented in this work not only deliver a practical solution for developing stable high-field (12-16 T) $Nb_3Sn$ magnets, but also extend beyond $Nb_3Sn$ systems, demonstrating broader applicability in enhancing the stability of various high-field superconducting magnets to address rapid localized quench events, such as high temperature superconducting magnets wound with second-generation (2G) coated conductor tapes. The established framework not only enables electromagnetic optimization across diverse engineering implementations but also stimulates further experimental and theoretical investigations into multiscale field-coupled stability problems. More fundamentally, the derived stability criteria and control philosophy shed new insights for managing instability phenomena in other complex multi-physics systems involving strong electromagnetic-thermal coupling.

**Acknowledgement**

C.X. and H.X.R acknowledge support from the National Natural Science Foundation of China (Grant Nos. 12372210 and 11972298).